# Initial Experience of Metabolic Imaging with Hyperpolarized [1-$^{13}$C]pyruvate MRI in Kidney Transplant Patients

Xiaoxi Liu[1], Ph.D., Ying-Chieh Lai[1,2], M.D., Di Cui[1], Ph.D., Shiang-Cheng Kung[3], M.D., Meyeon Park[3], M.D., MAS, Laszik Zoltan[4], M.D., Ph.D., Peder E.Z. Larson[1,*], Ph.D., Zhen J. Wang[1,*], M.D.

[1]Department of Radiology and Biomedical Imaging, University of California, San Francisco, San Francisco, California, USA

[2]Department of Medical Imaging and Intervention, Chang Gung Memorial Hospital at Linkou, Taoyuan 333, Taiwan

[3]Department of Medicine, University of California San Francisco Medical Center, San Francisco, California, USA

[4]Department of Pathology, University of California San Francisco Medical Center, San Francisco, California, USA

[*]**Corresponding Author:**

Zhen J. Wang

505 Parnassus Avenue, Room S-353

San Francisco, CA 94143

zhen.wang@ucsf.edu

Peder E.Z. Larson

1700 4$^{th}$ St

Byers Hall Room 102C

San Francisco, CA, USA 94143

peder.larson@ucsf.edu

## Grant Support

This work was supported by NIH grants P41EB013598, R21DK130002, R01CA249909 and S10OD025077.

## Running Title

Metabolic Imaging in Transplanted Kidneys

## Abstract

BACKGROUND: Kidney transplant is the treatment of choice for patients with end-stage renal disease. Early detection of allograft injury is important to delay or prevent irreversible damage.

PURPOSE: To investigate the feasibility of hyperpolarized (HP) [1-$^{13}$C]pyruvate MRI for assessing kidney allograft metabolism.

STUDY TYPE: Prospective.

SUBJECTS: 6 participants (mean age, 45.2 ± 12.4 years, 2 females) scheduled for kidney allograft biopsy and 5 patients (mean age, 59.6 ± 10.4 years, 2 females) with renal cell carcinoma (RCC).

FIELD STRENGTH/SEQUENCE: 3 Tesla, T2-weighted fast spin echo, multi-echo gradient echo, single shot diffusion-weighted echo-planar imaging, and time-resolved HP $^{13}$C metabolite-selective imaging.

ASSESSMENT: Five of the six kidney allograft participants underwent biopsy after MRI. Estimated glomerular filtration rate (eGFR) and urine protein-to-creatine ratio (uPCR) were collected within 4 weeks of MRI. Kidney metabolism was quantified from HP [1-$^{13}$C]pyruvate MRI using the lactate-to-pyruvate ratio in allograft kidneys and non-tumor bearing kidneys from RCC patients.

STATISTICAL TESTS: Descriptive statistics (mean ± standard deviation).

RESULTS: Biopsy was performed a mean of 9 days (range 5-19 days) after HP [1-$^{13}$C]pyruvate MRI. Three biopsies were normal, one showed low-grade fibrosis and one showed moderate microvascular inflammation. All had stable functioning allografts with eGFR > 60 mL/min/1.73 m$^2$ and normal uPCR. One participant who did not undergo biopsy had reduced eGFR of 49 mL/min/1.73 m$^2$ and elevated uPCR. The mean lactate-to-pyruvate ratio was 0.373 in participants with normal findings (n = 3) and 0.552 in participants with abnormal findings (n = 2). The lactate-to-pyruvate ratio was highest (0.847) in the participant with reduced eGFR and elevated uPRC. Native non-tumor bearing kidneys had a mean lactate-to-pyruvate ratio of 0.309.

DATA CONCLUSION: Stable allografts with normal findings at biopsy showed lactate-to-pyruvate ratios similar to native non-tumor bearing kidneys, whereas allografts with abnormal findings showed higher lactate-to-pyruvate ratios.

**Introduction**

Chronic kidney disease affects approximately 14% of American adults[1], and when it advances to end-stage renal disease is most cost-effectively treated by kidney transplantation[2]. Chronic allograft injury remains one of the biggest challenges in kidney transplantation, resulting in 20-30% of the allografts failing by 10 years[3]. Current methods for monitoring kidney allografts have important limitations. Noninvasive measurements such as serum creatinine / estimated glomerular filtration rate (eGFR) are known to have poor predictive value for allograft injury[4,5]. Therefore, biopsy remains the standard for assessing kidney allograft pathology, either via surveillance or indication biopsies[6]. Given the invasive nature of biopsy, surveillance biopsy is controversial in patients with stable kidney function. When declining kidney function as measured by a rise in serum creatinine triggers an indication biopsy, the degree of injury may already be advanced and irreversible[7]. Therefore, improved noninvasive methods for monitoring kidney allograft are needed to allow early identification of subclinical injury in order to guide appropriate biopsies and permit timely intervention.

Several imaging methods have been investigated in patients with kidney disease. Quantitative measurements of apparent diffusion coefficient (ADC) and $R_2^*$ relaxation rate have shown correlation with eGFR in patients[8,9]. However, most studies show overlap in measurements between normal and diseased state, and it remains to be determined whether these methods can monitor kidney injury prior to a decline in eGFR.

Altered metabolism may play an important role in various kidney diseases[10]. Of particular relevance is the metabolism of pyruvate, a critical substrate for energy metabolism[11]. It is metabolized to downstream products including lactate in the cytosol, and it is also shuttled into the mitochondria and converted to acetyl-CoA via the enzyme pyruvate dehydrogenase (PDH) in the first step of oxidative phosphorylation[11] The byproduct of this first step, $CO_2$, rapidly equilibrates with bicarbonate[11]. A shift of pyruvate metabolism from oxidative phosphorylation to glycolysis and lactate production has been noted in acute kidney injury[12,13] and chronic kidney disease such as diabetic nephropathy[10,14–16]. Hyperpolarized (HP) $^{13}C$ MRI is a molecular imaging method that allows rapid, noninvasive, pathway-specific investigation of dynamic metabolic processes[17]. In previous preclinical studies[16,18,19], HP $^{13}C$ pyruvate MRI has shown the potential to noninvasively monitor the shift to anaerobic glycolysis in injured kidneys with increased oxidative stress, inflammation, and tubular damage. Thus the aim of this study was to present initial experience applying HP [1-$^{13}C$]pyruvate MRI to assess metabolism in patients with kidney allograft.

## Materials and Methods

This study was approved by the Institutional Review Board. Written informed consent was obtained from all participants.

### Study Participants

Inclusion criteria for the study are: subjects 18 years or older who are scheduled to undergo either: 1) 6-month surveillance biopsy, or 2) indication biopsy because of allograft dysfunction. Exclusion criteria are: any subjects with contra-indication to MRI such as cardiac pacemaker. Six participants (mean age, 45.2 ± 12.4 years, 2 females) with kidney allograft who were scheduled for allograft biopsy were recruited. In addition, data from the non-tumor bearing native kidneys in five patients (mean age, 59.6 ± 10.4 years, 2 females) with renal cell carcinoma (RCC) who were enrolled in a separate IRB-approved HP [1-$^{13}$C]pyruvate MRI study[20] were used for comparison with data from kidney allografts.

## HP [1-$^{13}$C]pyruvate preparation and injection

HP $^{13}$C pyruvate injection was approved under an Investigational New Drug (IND157192) protocol by the US Food and Drug Administration. To prepare HP $^{13}$C pyruvate, an aliquot of 1.47g [1-$^{13}$C]pyruvate acid (ISOTEC, MilliporeSigma, Miamisburg, OH, USA) and 15 mM trityl electron paramagnetic agent (AH111501; GE HealthCare, Chicago, IL, USA) were mixed and polarized using a SPINlab polarizer (GE Research, Niskayuna, New York, USA) operating at 5T and 0.8 Kelvin for at least 2.5 hours and no more than 4 hours. Following dissolution with sterile water, filtration by a mechanical filter (0.2 μm, ZenPure, VA, USA), neutralization by equivalent NaOH and Tris aminomethane and release by a pharmacist, the HP [1-$^{13}$C]pyruvate agent was injected intravenously into the study participant at a 0.43 mL/kg dose followed by a 20 mL saline flush at 5 mL/s. The $^{13}$C scan started at the end of saline flush, and the participant breathed freely during the HP $^{13}$C acquisition.

## MRI Protocol

All studies were performed on a 3 Tesla GE scanner (MR750, GE Healthcare, Waukesha, WI, USA) equipped with clinical performance gradients (gradient$_{max}$ = 50 mT/m, slew-rate$_{max}$ = 200 T/m/s), and included $^1$H MRI and HP $^{13}$C MRI acquisitions.

The coil setup and HP $^{13}$C MRI protocol are shown in Figure 1. The HP $^{13}$C MR protocol was implemented on a commercial software platform (RTHawk Research, Vista.ai Inc., Los Altos, CA, USA), and HP $^{13}$C data were acquired using a semi-flexible quadrature transmit and 8-channel receive array coil (QTAR, Clinical MR Solutions, Brookfield, WI, USA). The kidney allograft was centrally positioned within the coil to ensure complete coverage and the positioning was confirmed using fiducial markers located at the edge of the coil by a 2D gradient echo (GRE) sequence. At the end of the saline flush, HP $^{13}$C imaging started with a real-time prescan protocol[21,22] including bolus detection, B$_0$ calibration and B$_1$ calibration, as shown in Figure 1C. Following the detection of signal-to-noise ratio (SNR) over the threshold (2× noise level), the B$_0$ calibration on the [1-$^{13}$C]pyruvate signal frequency, B$_1$ calibration and dynamic HP $^{13}$C metabolite imaging on the kidney allograft were automatically triggered. The total acquisition time of the HP $^{13}$C protocol was approximately 2 minutes. Scan parameters were: 50 cm field-of-view (FOV); 1 cm in-plane resolution for pyruvate-specific 2D GRE sequence (80 ms TR) and lactate-specific 3D balanced steady-state free precession (bSSFP) sequence (15.3 ms TR)[23]; 2 cm in-plane resolution for bicarbonate-specific 3D bSSFP sequence (9.8 ms TR)[24]; 21 mm slice thickness; 20° flip angle for pyruvate, 60° flip angle for lactate and bicarbonate; 3.5s temporal resolution. In one participant with renal allograft, two repeated injections separated by approximately 20 minutes were performed. In the second HP $^{13}$C injection, $^{13}$C images were acquired at a lower resolution: 1.5 cm in-plane resolution for pyruvate and lactate, 2.5 cm in-plane resolution for bicarbonate, while the other acquisition parameters were the same.

The $^1$H MRI protocol included anatomical reference images as well as data for shimming, diffusion weighted image (DWI), and $T_2^*$. Anatomic images were acquired before the HP $^{13}$C injection using a multi-slice single-shot $T_2$-weighted fast spin echo (FSE) sequence (TR/TE = 3500/85.1 ms, FOV = 420 × 420 mm$^2$, resolution = 1.64 × 1.64 × 6 mm$^3$). Additionally, a shimming calibration was performed using a 3D multi-echo gradient echo (IDEAL-IQ) sequence (TR/TE = 5.8/2.6 ms, echo train length = 3, FOV = 440 × 440 mm$^2$, resolution = 2.75 × 2.75 × 10 mm$^3$) to pre-set shim values for the HP $^{13}$C MR scan. This multi-echo data was also used for $T_2^*$ measurements. Following the HP $^{13}$C data acquisition, diffusion-weighted imaging was acquired using single-shot echo planar imaging sequence (TR/TE = 4884/66.9 ms, FOV = 420 × 420 mm$^2$, resolution = 4.38 × 6.56 × 6 mm$^3$, b values = 25, 50, 75, 100, 300, 600, 800, 1000 s/mm$^2$).

No change in vital signs and no adverse effects were observed following the HP $^{13}$C pyruvate MRI in any of the study participants.

**Data Analysis**

HP $^{13}$C data were reconstructed, gridded and inverse Fourier transformed to the image domain[25] on MATLAB 2019a platform (Mathworks Inc., Natick, MA, USA). Coil sensitivity maps were estimated by area under the curve (AUC) data of multi-channel pyruvate signal and then were used for channel combination[26] of all dynamic metabolite signals after noise pre-whitening[27]. Images were zero-filled with a 2D Fermi filter to match the resolution of 2D FSE images, which served as an anatomical reference for the following quantitative assessment.

Regions-of-interest (ROIs) in the cortex of kidney allografts and of the native kidneys from the RCC study were manually drawn on the $^{13}$C metabolite maps by an imaging research fellow (XL, 2 years of experience in kidney imaging research) in consultation with a fellowship-trained abdominal radiologist (ZW, 16 years of experience) using T2-weighted FSE images as a reference.

To quantify metabolism, normalized lactate-to-pyruvate and bicarbonate-to-pyruvate ratios[28] were calculated from lactate or bicarbonate AUC values as following: 1) normalized lactate-to-pyruvate ratio = lactate AUC summed over all time points / maximum pyruvate AUC; 2) normalized bicarbonate-to-pyruvate ratio = bicarbonate AUC summed over all time points / maximum pyruvate AUC. The bicarbonate AUC values were also normalized based on the acquired voxel size ratio by dividing by (2 cm × 2 cm) / (1 cm × 1 cm) = 4, since bicarbonate was acquired with 2 cm in-plane resolution versus 1 cm for the other metabolites. Finally, the AUC values were also normalized by noise level measured in the last five timepoints. The acquisition and bolus timing can also have an impact and introduce errors in the AUC analysis. To normalize for the timing of the dynamic acquisition, the center of mass or "mean time"[29] of the pyruvate dynamic signal curve was made consistent across studies by removing some initial timepoints (either 3 or 4) prior to the AUC calculations[30].

ADC maps from $^1$H diffusion weighted imaging were processed using MRtrix3 software[31]. $R_2^*$ maps from IDEAL-IQ sequence were calculated and generated by the built-in software on the GE scanner.

**Histopathology and laboratory evaluation**

Five of the six participants underwent allograft biopsy following MRI. The histopathological evaluation of the biopsy tissues was performed as per clinical practice using the updated Banff 2019 criteria[32], and biopsy findings were recorded. For all 6 participants with renal allograft, estimated glomerular filtration rate (eGFR) and urine protein-to-creatine ratio (uPCR) closest to the MRI date and within 4 weeks of the MRI was also recorded. In addition, for all RCC patients, the eGFR within 4 weeks of and closest to the MRI date was also recorded.

**Results**

Data from six kidney allografts and five native non-tumor bearing kidneys in patients with RCC in a separate study were included. A summary of the study participants' demographics is shown in Table 1. Five of the six participants with kidney allografts underwent allograft biopsy a mean of 9 days (range 5-19 days) following HP [1-$^{13}$C]pyruvate MRI. One participant did not undergo the scheduled allograft biopsy due to Covid-19 infection.

The bolus was detected in the allograft 19.83 ± 2.55 seconds after the start of the injection. The bolus was detected in the native kidney 18.80 ± 2.86 seconds after the start of the injection, which is consistent with previously reported time range of bolus arrival in the native kidney (15 to 28 seconds)[20]. The mean +/- standard deviation of the AUC SNR for pyruvate, lactate and bicarbonate were 37.91 ± 22.27, 20.60 ± 10.28, and 9.60 ± 3.57.

Figure 2 shows dynamic metabolite images through two representative slices of two renal allografts. The dynamic images show the delivery of $^{13}$C pyruvate to the allografts followed by metabolic conversion to $^{13}$C-lactate and $^{13}$C-bicarbonate. Due to inhomogeneity caused by the RF coil profiles, the metabolite signals were higher in the more anterior aspect of the kidney allografts which are closer to the receive coil elements. This inhomogeneity is accounted for when normalizing to the $^{13}$C pyruvate AUC in the AUC ratio analyses. The summation of these dynamic metabolite images over time and encompassing the entire allograft was used for metabolite quantification.

Table 1 shows metabolite AUC ratios of kidney allografts and native non-tumor bearing kidneys in patients with RCC from a separate study, allograft biopsy findings, eGFR and uPCR within 4 weeks of MRI for participants with kidney allograft. Of the five participants who underwent kidney allograft biopsy, three had normal findings on biopsy, one had allograft showing moderate microvascular inflammation, and one had allograft showing low grade interstitial fibrosis and tubular atrophy. All five participants had stable functioning allografts with eGFR > 60 mL/min/1.73 m$^2$ and with normal range of uPCR. The participant who did not undergo biopsy after MRI due to Covid-19 infection had reduced eGFR of 49 mL/min/1.73 m$^2$ and elevated uPCR of 4462. The mean lactate-to-pyruvate ratio was 0.373 (range, 0.319-0.420) in the three participants with allografts that had normal findings at biopsy, which was lower than the mean of 0.552 (range, 0.542-0.561) in the two participants with abnormal findings at biopsy. The lactate-to-pyruvate ratio was the highest at 0.847 in the participant who did not have post-MRI biopsy. The bicarbonate-to-pyruvate ratios were variable across the kidney allografts. Figure 3 shows the $^{13}$C pyruvate, lactate and lactate-to-pyruvate AUC ratio images of the six kidney allografts. A summary of all metabolite AUC images is shown in Figure S1.

The mean lactate-to-pyruvate ratio of the five native non-tumor bearing kidneys was 0.309 (range, 0.248-0.389), similar to that of the two allografts with normal biopsy findings. All five RCC patients had eGFR > 60 mL/min/1.73 m$^2$ within 4 weeks of their MRI. Figure 4a shows the distribution of mean values of lactate-to-pyruvate ratio of kidney allografts and native kidneys.

Table S1 includes the ADC and $R_2^*$ values measured in kidney allografts and native kidneys. The ADC values of allografts with normal biopsy findings are similar to those of native kidneys. The ADC values of the allograft with low grade fibrosis and of the allograft with reduced eGFR were lower than those of allografts with normal biopsy findings and native kidneys. The $R_2^*$ values were variable across the allografts and the native kidneys. Figure 4b and 4c show the distribution of ADC and $R_2^*$ of kidney allografts and native kidneys. Figures S2 and S3 show representative ADC and $R_2^*$ maps of kidney allografts.

The metabolite AUC images of the one patient who had two repeated HP [1-$^{13}$C]pyruvate injections approximately 20 minutes apart are shown in Figure 5. With a higher spatial resolution, the images from the first injection show sharper features but with a lower SNR.

Images from the repeated injections show similar lactate-to-pyruvate (Injection 1: 0.38 ± 0.03, and Injection 2: 0.41 ± 0.02) and bicarbonate-to-pyruvate (Injection 1: 0.040 ± 0.003, and Injection 2: 0.047 ± 0.002) AUC ratios, suggesting the potential reproducibility of the technique for quantifying metabolite ratios.

Table S2 shows comparison of metabolite ratios with "mean time" correction and without "mean time" correction. The "mean time" correction was important to account for the variability in timing of the dynamic acquisition.

## Discussion

In this study, initial data on HP [1-$^{13}$C]pyruvate MRI for assessing metabolism in kidney allografts were presented. The three stable functioning allografts with normal findings at biopsy had lactate-to-pyruvate ratios similar to those of native non-tumor bearing kidneys in patients with RCC from a separate study, and the two allografts with abnormal findings at biopsy had higher lactate-to-pyruvate ratios compared to allografts with normal findings at biopsy.

The kidney is a highly energy dependent organ requiring mitochondrial oxidative phosphorylation for key renal functions and protection against injuries[11]. Given that pyruvate is a critical substrate for energy metabolism, it is important to establish the normal range of pyruvate metabolism in well-functioning kidneys in order to better understand altered metabolism associated with kidney injury. Our initial data showed that stable functioning allografts with normal findings at biopsy, an eGFR > 60 (mL/min/1.73m$^2$), and lack of proteinuria (normal range of uPCR) had lactate-to-pyruvate ratios in the same range as those from well-functioning native kidneys. A previous study has shown that patients with kidney allograft whose eGFR was > 60 (mL/min/1.73m$^2$) and without proteinuria had the lowest risk of progressive allograft injury and allograft loss[33] ; that study supports the current study which showed that the observed metabolism from pyruvate to lactate in this group reflected that of a well-functioning kidney, and the normal biopsy findings in this group also supports allograft health.

Several prior preclinical studies have investigated HP pyruvate metabolism in murine models of kidney injury. Elevated renal lactate-to-pyruvate ratios were observed in models of diabetes[16], ureteral obstruction[19] and ischemia-reperfusion injury[18], and were associated with increased inflammation in the ischemic kidneys, and with increased inflammation and fibrosis in the obstructed kidneys. The higher lactate-to-pyruvate ratios in the two allografts with abnormal biopsy findings (moderate microvascular inflammation, low grade fibrosis) in the current study are in agreement with findings from the preclinical studies and suggest the potential of HP pyruvate MRI for detecting altered metabolism associated with kidney injury in patients. It is also interesting to note that the two participants with abnormal allograft biopsy findings had stable functioning allografts with eGFR > 60 (mL/min/1.73 m$^2$) and no proteinuria at the time of the MRI, while their allografts showed higher lactate-to-pyruvate ratio compared to the allografts with normal biopsy findings. It is recognized that although monitoring eGFR and urine protein is the accepted approach for assessing allograft function, these tests are not sensitive for detecting subclinical injury[34]. Larger studies are needed to investigate whether metabolic alterations observed in HP pyruvate MRI precede a decrease in eGFR and the appearance of proteinuria and whether the technique can be used for early noninvasive detection of allograft injury. The highest kidney allograft lactate-to-pyruvate ratio was observed in the participant with reduced eGFR and proteinuria. Of note, this participant underwent allograft biopsy 7 months prior to MRI (2-days post-transplant) which showed acute tubular injury and low-grade tubulitis. While this participant did not undergo the subsequent planned allograft biopsy due to Covid-19 infection, the finding of elevated lactate-to-pyruvate ratio was consistent with our observations in the five participants with biopsy results and this participant's laboratory findings likely indicate underlying allograft injury.

The bicarbonate-to-pyruvate ratios were variable across the kidney allografts in our study. A prior preclinical study showed reduced renal bicarbonate-to-pyruvate ratio in a murine model of ischemia reperfusion injury due to reduced PDH activity[18]. Another study showed no change in renal bicarbonate-to-pyruvate ratios in a murine model of diabetes, and it was hypothesized that this was due to the occurrence of pseudohypoxia in diabetes where sufficient oxygen is supplied to the kidney to maintain PDH activity[16]. It is possible that the kidney PDH activity can vary depending on the type of kidney injury / disease, and further studies are required to investigate whether HP pyruvate MRI can detect the alteration in PDH activity.

The most common cause of kidney allograft failure after the first year is chronic allograft nephropathy, which is a histopathology description that refers to the features of interstitial fibrosis and tubular atrophy[35]. It has been shown that impaired mitochondrial energy metabolism leads to increased inflammation in the kidneys[14], and chronic inflammation is thought to be a mediator of kidney allograft interstitial fibrosis and progressive injury[12]. Molecular profiling of surveillance biopsy tissues has also demonstrated a signature of impaired mitochondrial energy metabolism in stable functioning allografts that show histological findings of inflammatory fibrosis[36]. Chronic allograft nephropathy can be a result of immunological and nonimmunological etiologies including rejection, ischemia, medication (calcineurin inhibitor) toxicity, infection, and recurrent disease[37]. While many of these etiologies can lead to altered metabolism which can potentially be detected using HP pyruvate MRI, the observed metabolic alterations likely cannot reliably differentiate among the different etiologies that result in chronic allograft nephropathy. Nonetheless, monitoring kidney allograft pyruvate metabolism may improve the early detection of injury and thereby guide appropriate biopsy.

Several functional imaging methods have been investigated in kidney diseases. ADC from DWI and $R_2^*$ from blood oxygen level dependent (BOLD) MRI have shown correlation to eGFR in patients with kidney allografts[8,9]. It remains to be determined whether these measurements can monitor allograft injury prior to a decline in eGFR. In the current study, the ADC values of the allograft with low grade fibrosis and of the allograft with reduced eGFR and with proteinuria were lower than those of allografts with normal biopsy findings and native kidneys, suggesting the potential utility of including DWI for monitoring kidney allografts. The $R_2^*$ values were variable across the allografts and the native kidneys. Future studies are needed to investigate whether multiparametric evaluation incorporating both HP metabolic data and $^1$H data can improve early detection of allograft injury. As HP $^{13}$C pyruvate MR images are acquired rapidly (within 1-2 minutes), this can be readily incorporated into a multiparametric MRI examination.

In the current study, pyruvate bolus tracking and real-time $B_0/B_1^+$ calibration methods[22] were used for HP pyruvate acquisition; this is expected to improve the robustness and reproducibility of metabolism quantifications. As the pyruvate arrival time in the kidneys can vary among subjects (for example, due to differences in cardiac output and the patency of the reconstructed renal artery) and the imaging window is short for the HP $^{13}$C signal (~1 min), the bolus tracking method can provide more consistent timing of the HP $^{13}$C MR acquisition. The real-time $B_0/B_1^+$ calibrations have the potential to provide consistency and knowledge of the actual imaging flip angles, which may improve both image quality and the reproducibility of the metabolite AUCs and metabolite AUC SNR ratios. For one of the patients with renal allograft who had two HP pyruvate injections followed by MRI, the lactate-to-pyruvate and bicarbonate-to-pyruvate AUC ratios were similar between the two acquisitions. While this suggests that the technique may be reproducible, much larger studies are needed to confirm this initial observation.

## Limitations

First, the number of study participants was small in this pilot study. Nonetheless we observed differences in lactate-to-pyruvate ratios in allografts with normal versus abnormal biopsy findings. Although the small numbers precluded statistical analysis, these initial observations provide rationale for conducting larger studies in kidney allografts to investigate the utility of HP pyruvate MRI for detecting early allograft injury. Second, we compared the allograft metabolism to that of non-tumor bearing native kidneys in RCC patients from a separate study that was published before[20]. It is possible that metabolism of the non-tumor bearing native kidney is affected by the presence of tumor in the contralateral kidney. Future studies in healthy volunteers are needed to assess normal native kidney pyruvate metabolism and whether kidney allograft metabolism differs from native kidney metabolism. There are other reported HP pyruvate MRI data in non-tumor bearing kidneys of RCC patients in the literature such as those by Ursprung et al [38], where the reported lactate-to-pyruvate ratio was 0.14 (0.12 - 0.22) for the non-tumor bearing kidneys, lower than those observed in this study. However, that study used a different (chemical shift imaging) acquisition technique with different flip angles and timings that affect the lactate-to-pyruvate ratio. Pharmacokinetic modeling of HP pyruvate MRI has the potential to more readily enable comparison of data across studies, and we are actively developing and exploring such modeling for use in future analyses [28]. Third, due to the available SNR of each metabolite, the spatial resolution was relatively coarse. As such, there is likely partial volume effect which limits the distinction between renal cortical and medullary metabolism. A variable resolution acquisition would enable higher resolution for pyruvate and lactate and reduce partial volume effects[39], and will be explored in future studies. Fourth, we did not consider the $T_2^*$ variation[40] of pyruvate during acquisition. Additional $T_2^*$ correction can lead to a more accurate metabolism quantification and will be investigated in future studies.

**Conclusion**

HP [1-$^{13}$C]pyruvate MRI in kidney transplant patients showed metabolic conversion to lactate and bicarbonate with excellent SNR. Stable functioning kidney allografts with normal findings at biopsy showed lactate-to-pyruvate ratios similar to those of well-functioning native kidneys. Allografts with abnormal findings at biopsy showed elevated lactate-to-pyruvate ratios compared to those from allografts with normal findings at biopsy.

**Acknowledgement**

We thank Robert Bok, Mary Frost, Heather Daniel, Kimberly Okamoto, Duy Dang, Evelyn Escobar, Stacy Andosca, Hsin-Yu Chen, and Romelyn Delos Santos for assistance with the human study.

21# References

1. Kidney Disease Statistics for the United States. Accessed December 9, 2023. https://www.niddk.nih.gov/health-information/health-statistics/kidney-disease

2. Einecke G, Broderick G, Sis B, Halloran PF. Early loss of renal transcripts in kidney allografts: Relationship to the development of histologic lesions and alloimmune effector mechanisms. *American Journal of Transplantation*. 2007;7(5):1121-1130. doi:10.1111/j.1600-6143.2007.01797.x

3. Callus R, Takou A, Sharma A, Halawa A. *Chronic Allograft Dysfunction in the Renal Transplant Recipient: An Ongoing Challenge for the Transplant Physician*. Vol 1.; 2019. www.scitcentral.com

4. Alaini A, Malhotra D, Rondon-Berrios H, et al. Establishing the presence or absence of chronic kidney disease: Uses and limitations of formulas estimating the glomerular filtration rate. *World J Methodol*. 2017;7(3):73-92. doi:10.5662/wjm.v7.i3.73

5. Bidin MZ, Shah AM, Stanslas J, Lim CTS. Blood and urine biomarkers in chronic kidney disease: An update. *Clinica Chimica Acta*. 2019;495:239-250. doi:10.1016/j.cca.2019.04.069

6. Moein M, Papa S, Ortiz N, Saidi R. Protocol Biopsy After Kidney Transplant: Clinical Application and Efficacy to Detect Allograft Rejection. *Cureus*. Published online February 1, 2023. doi:10.7759/cureus.34505

7. Kellum JA, Romagnani P, Ashuntantang G, Ronco C, Zarbock A, Anders HJ. Acute kidney injury. *Nat Rev Dis Primers*. 2021;7(1). doi:10.1038/s41572-021-00284-z

8. Li LP, Milani B, Pruijm M, et al. Renal BOLD MRI in patients with chronic kidney disease: comparison of the semi-automated twelve layer concentric objects (TLCO) and manual ROI methods. *Magnetic Resonance Materials in Physics, Biology and Medicine*. 2020;33(1):113-120. doi:10.1007/s10334-019-00808-5

9. Yalçin-Şafak K, Ayyildiz M, Ünel SY, Umarusman-Tanju N, Akça A, Baysal T. The relationship of ADC values of renal parenchyma with CKD stage and serum creatinine levels. *Eur J Radiol Open*. 2016;3:8-11. doi:10.1016/j.ejro.2015.10.002

10. Zepeda-Orozco D, Kong M, Scheuermann RH. Molecular Profile of Mitochondrial Dysfunction in Kidney Transplant Biopsies Is Associated with Poor Allograft Outcome. *Transplant Proc*. 2015;47(6):1675-1682. doi:10.1016/j.transproceed.2015.04.086

## Tables

**Table 1:** Summary of study participant demographics, clinical data, allograft biopsy findings, and metabolic area under the curve (AUC) ratios.

| Participants | Age (year) | Sex | Reason for Kidney Transplant | Time from Transplant to MRI (month) | Time from Imaging to Biopsy (day) | Clinicopathology | | | Lactate-to-Pyruvate Ratio | Bicarbonate-to-Pyruvate Ratio |
|---|---|---|---|---|---|---|---|---|---|---|
| | | | | | | eGFR* | uPCR* | Biopsy Findings | | |
| **Participants with kidney Allograft** | | | | | | | | | | |
| #1 | 52 | Male | Diabetic nephropathy | 6 | 9 | 67 | 121 | Normal | 0.42±0.04 | 0.05±0.01 |
| #2 | 48 | Female | Polycystic kidney disease | 5.5 | 19 | 72 | 84 | Normal | 0.38±0.03 | 0.04±0.00 |
| #3 | 40 | Female | Diabetic nephropathy | 5.5 | N/A | 46 | 4462 | N/A | 0.85±0.07 | 0.11±0.01 |
| #4 | 24 | Female | End-stage renal disease of unknown etiology | 6 | 7 | 95 | 105 | Moderate microvascular inflammation | 0.54±0.05 | N/A |
| #5 | 58 | Male | Diabetic nephropathy | 6 | 5 | 65 | 90 | Low grade fibrosis | 0.56±0.04 | 0.06±0.01 |
| #6 | 61 | Male | HIV nephropathy | 6 | 7 | 64 | 44 | Normal | 0.32±0.03 | 0.04±0.00 |
| **RCC patients with native non-tumor bearing kidneys** | | | | | | | | | | |
| #1 | 46 | Male | N/A | N/A | N/A | 83 | N/A | N/A | 0.39±0.03 | N/A |
| #2 | 61 | Male | N/A | N/A | N/A | 71 | N/A | N/A | 0.25±0.04 | N/A |
| #3 | 47 | Male | N/A | N/A | N/A | 103 | N/A | N/A | 0.25±0.02 | N/A |
| #4 | 54 | Female | N/A | N/A | N/A | 70 | N/A | N/A | 0.37±0.03 | N/A |
| #5 | 61 | Female | N/A | N/A | N/A | 73 | N/A | N/A | 0.39±0.02 | 0.06±0.01 |

Note: Values are mean ± standard deviation. eGFR: estimated glomerular filtration rate (mL/min/1.73 m$^2$), HIV: human immunodeficiency virus, N/A: Not available, RCC: renal cell carcinoma, uPCR: urine protein-to-creatinine ratio (normal range, < 150mg/g).

*; obtained within 4 weeks of MRI for patients with kidney allograft

**Figure Legends**

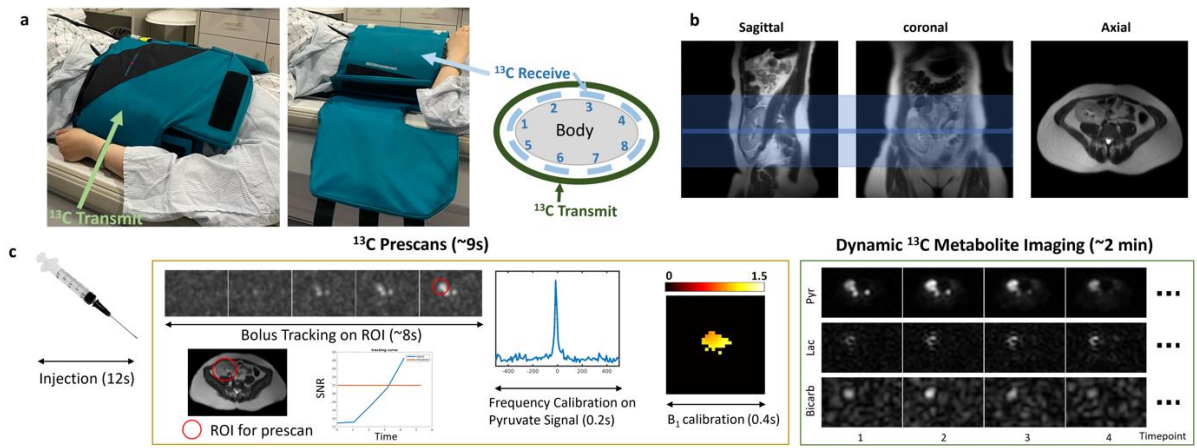

Figure 1. Coil setup and HP $^{13}$C study protocol. (a) positioning of the semi-flexible quadrature transmit and 8-channel receive array coil. (b) Localized $^1$H MRI scans acquired as an anatomical reference for the HP $^{13}$C scan. The blue region shows the coverage of the multi-slice HP $^{13}$C scan. (c) HP $^{13}$C MRI acquisition, which involves bolus detection and $B_0$/$B_1^+$ field calibrations in the real-time prescan and dynamic $^{13}$C metabolite-specific imaging.

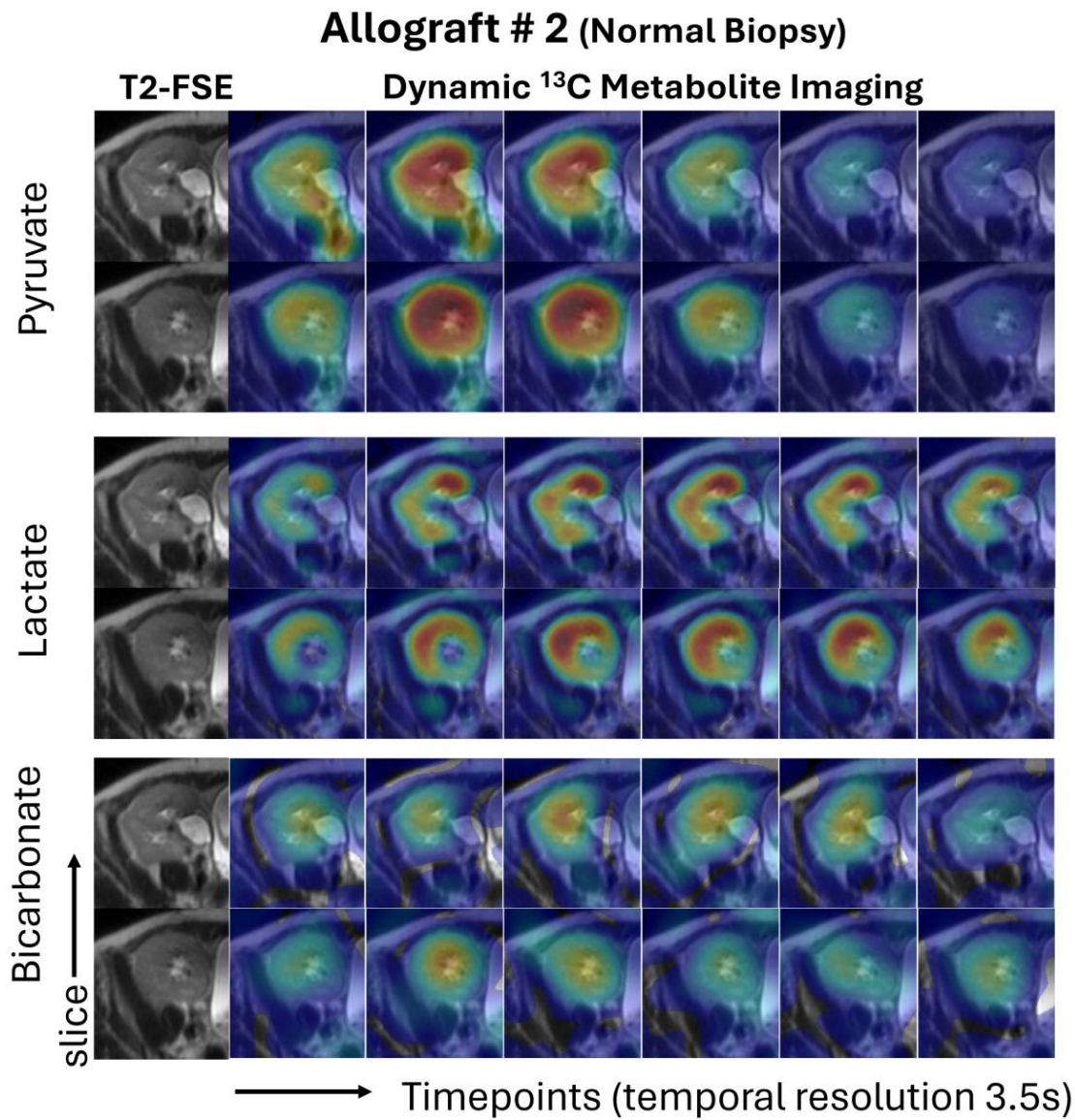

Figure 2. Dynamic metabolite images through two representative slices in a renal allograft with a normal biopsy. The dynamic images show the delivery of $^{13}$C pyruvate to the allograft (top) followed by metabolic conversion to $^{13}$C-lactate (mid) and $^{13}$C-bicarbonate (bottom). Due to inhomogeneity caused by the RF coil profile, the metabolite signals were higher in the more anterior aspect of the kidney allografts which are closer to the coil. This inhomogeneity is accounted for using the area under the curve (AUC) ratio analysis.

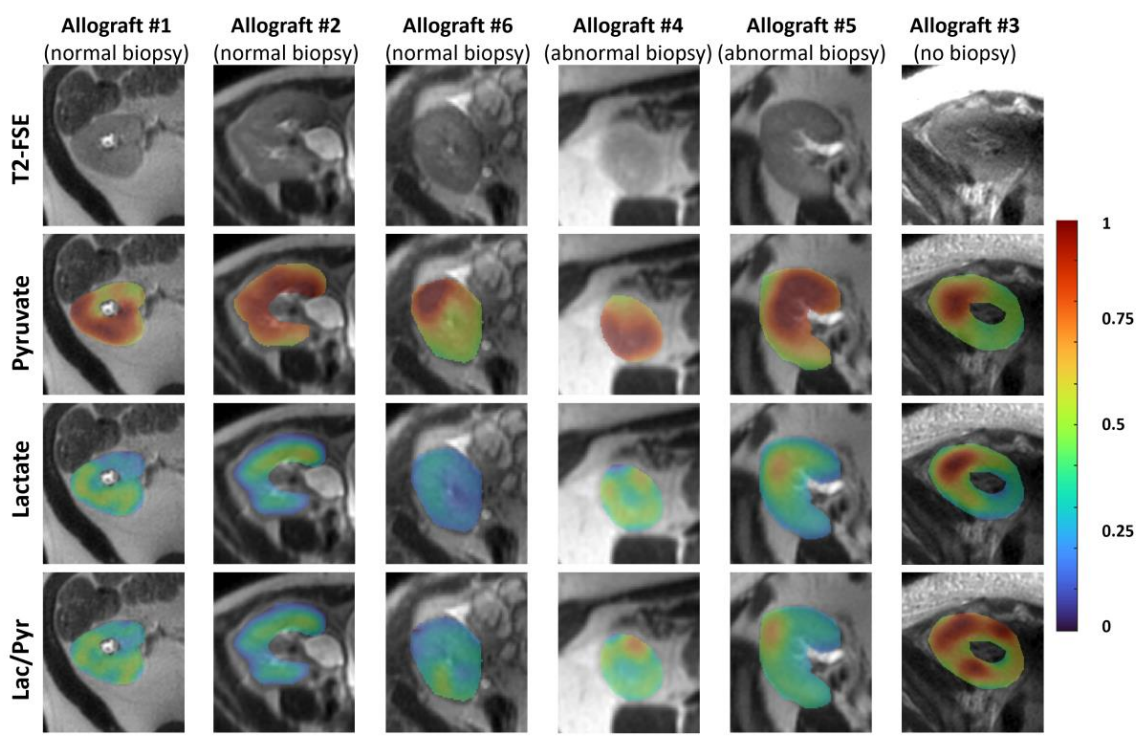

Figure 3. $^{13}$C pyruvate, lactate and lactate-to-pyruvate ratio images in six kidney allografts. All AUC images were normalized by the maximum kidney pyruvate AUC.

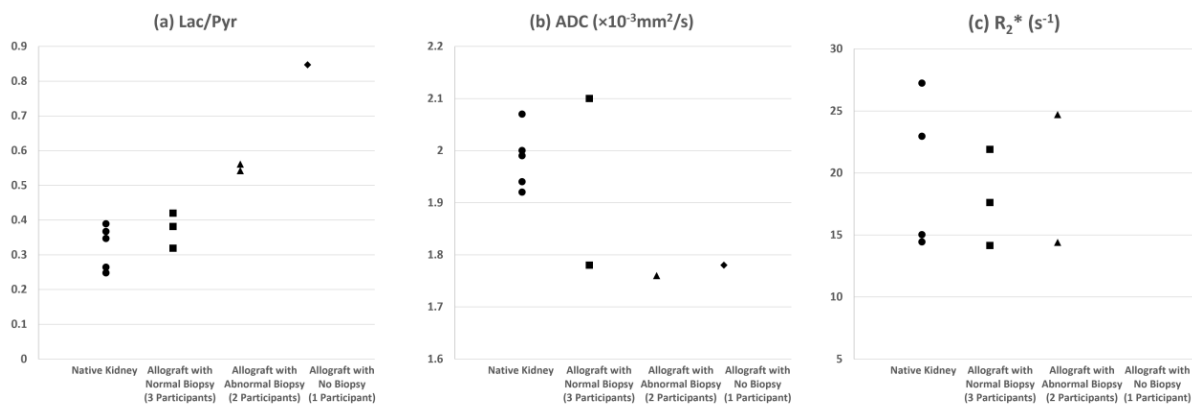

Figure 4. Distribution of mean values of (a) lactate-to-pyruvate ratio, (b) apparent diffusion coefficient (ADC) and (c) $R_2^*$ of kidney allografts and native non-tumor bearing kidneys in patients with renal cell carcinomas.

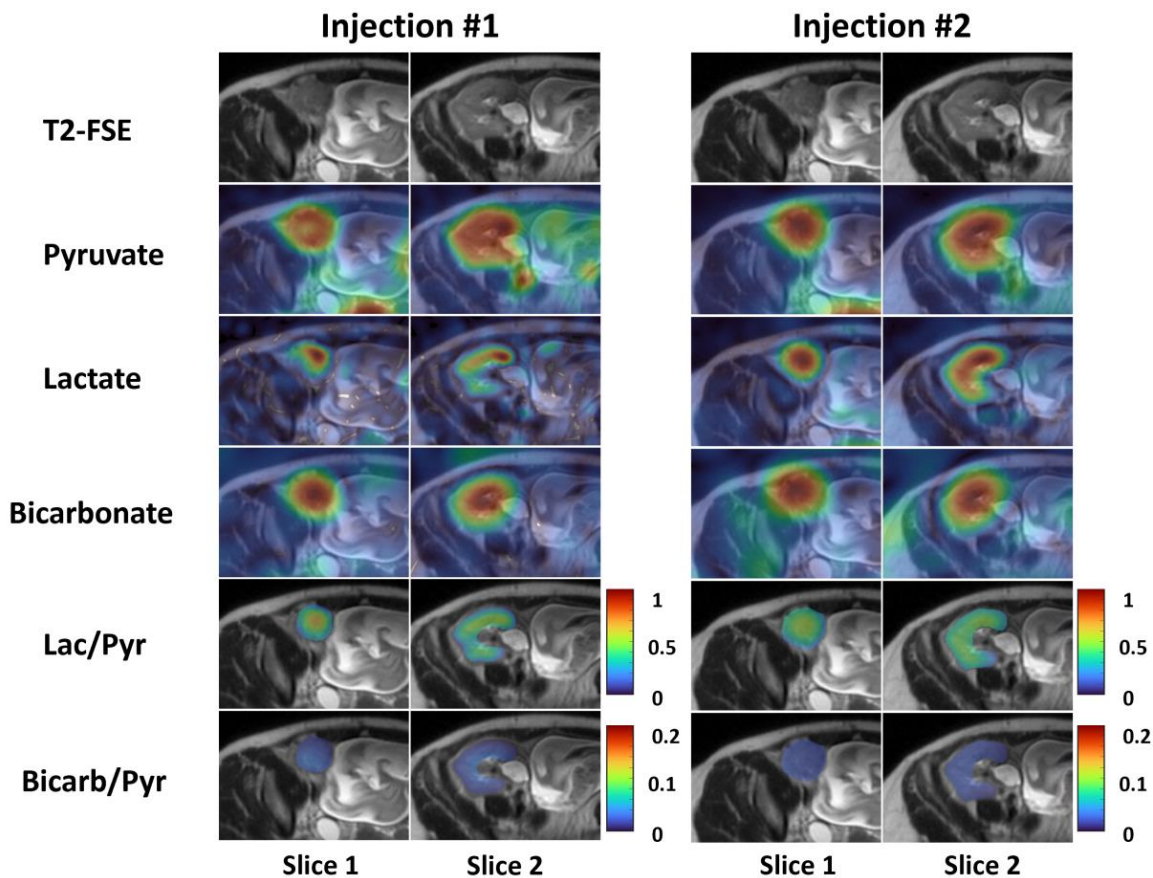

Figure 5. $^{13}$C AUC and metabolite ratio images in the patient with two repeated injections approximately 20 minutes apart. In the first injection, pyruvate and lactate images were acquired at 1 cm in-plane resolution, and bicarbonate images were acquired at 2 cm in-plane resolution. In the second injection, pyruvate and lactate images were acquired at 1.5 cm in-plane resolution, and bicarbonate images were acquired at 2.5 cm in-plane resolution. Images from the repeated injections show similar lactate-to-pyruvate and bicarbonate-to-pyruvate AUC ratios, suggesting the potential reproducibility of the technique for quantifying metabolite ratios.

# Supplementary Information

## Tables

**Table S1:** Apparent diffusion coefficient (ADC) and $R_2^*$ of kidney allografts and native non-tumor bearing kidneys from patients with renal cell carcinomas.

| Participant | ADC ($\times 10^{-3}$ mm$^2$/s) | $R_2^*$ (s$^{-1}$) |
|---|---|---|
| **Participants with kidney Allograft** | | |
| #1 | N/A | 21.89±14.59 |
| #2 | 2.10±0.11 | 14.15±9.21 |
| #3 | 1.78±0.26 | N/A |
| #4 | N/A | 14.41±8.62 |
| #5 | 1.76±0.08 | 24.70±14.71 |
| #6 | 1.92±0.09 | 17.61±10.16 |
| **RCC patients with native non-tumor bearing kidneys** | | |
| #1 | 1.99±0.09 | N/A |
| #2 | 1.94±0.10 | 15.03±8.06 |
| #3 | 2.07±0.10 | 27.24±12.77 |
| #4 | 2.00±0.09 | 22.96±9.81 |
| #5 | 1.92±0.09 | 14.44±9.63 |

Note: N/A: Not available, either due to not acquired (ADC) or insufficient signal-to-noise ratio (SNR) for quantification ($R_2^*$).

**Table S2:** Comparison of lactate-to-pyruvate ratio and bicarbonate-to-pyruvate ratio of kidney allografts and native non-tumor bearing kidneys with "mean time" correction and without "mean time" correction.

| Participant | Participants with kidney Allograft | | | |
|---|---|---|---|---|
| | With Mean Time Correction | | Without Mean Time Correction | |
| | Lactate-to-Pyruvate Ratio | Bicarbonate-to-Pyruvate Ratio | Lactate-to-Pyruvate Ratio | Bicarbonate-to-Pyruvate Ratio |
| #1 | 0.42±0.04 | 0.05±0.01 | 0.42±0.04 | 0.05±0.01 |
| #2 | 0.38±0.03 | 0.040±0.03 | 0.21±0.01 | 0.03±0.00 |
| #3 | 0.85±0.07 | 0.11±0.01 | 0.85±0.07 | 0.11±0.01 |
| #4 | 0.54±0.05 | N/A | 0.24±0.02 | N/A |
| #5 | 0.56±0.04 | 0.06±0.01 | 0.38±0.04 | 0.04±0.04 |
| #6 | 0.32±0.03 | 0.04±0.00 | 0.32±0.03 | 0.04±0.00 |
| RCC patients with native non-tumor bearing kidneys | | | | |
| #1 | 0.39±0.03 | N/A | 0.28±0.02 | N/A |
| #2 | 0.25±0.04 | N/A | 0.19±0.01 | N/A |
| #3 | 0.25±0.02 | N/A | 0.22±0.02 | N/A |
| #4 | 0.37±0.03 | N/A | 0.37±0.03 | N/A |
| #5 | 0.39±0.02 | 0.06±0.01 | 0.39±0.03 | 0.06±0.01 |

Note: N/A: Not available.

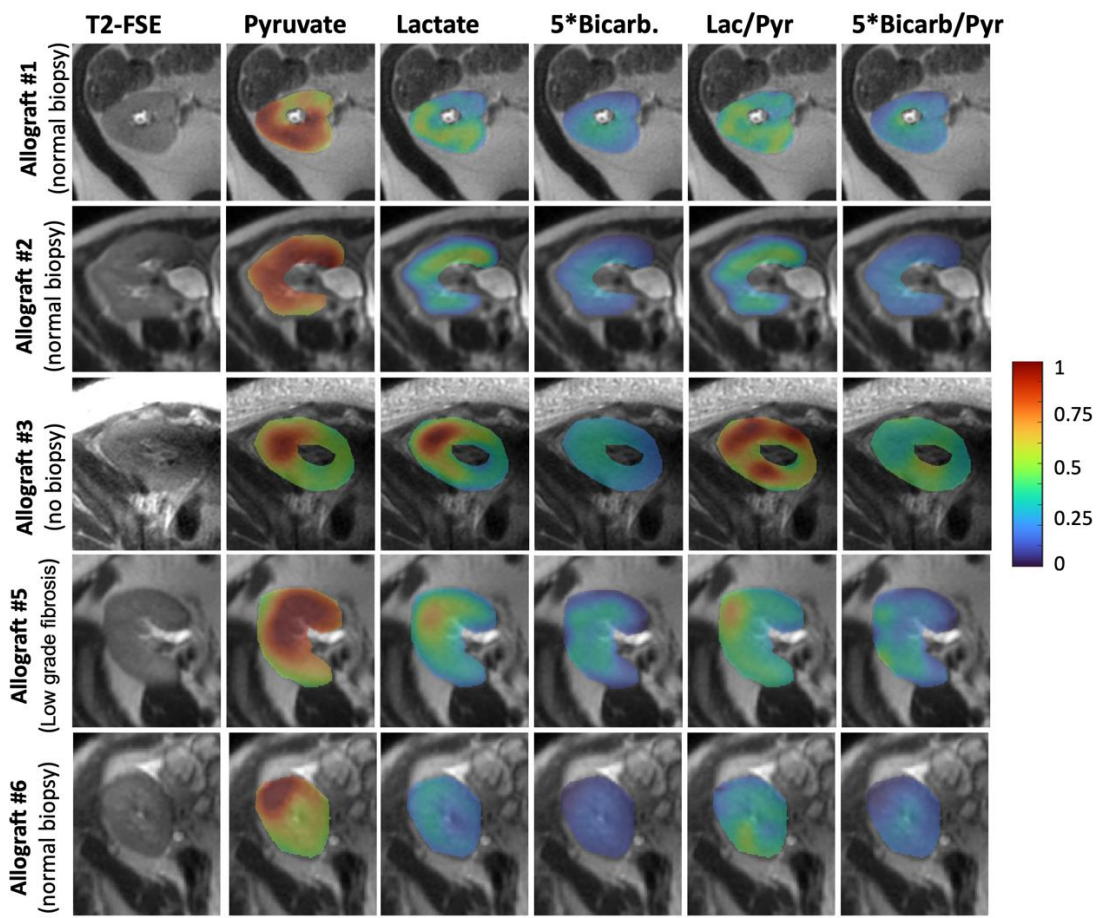

Figure S1 $^{13}$C pyruvate, lactate, bicarbonate and metabolite ratio images from five kidney transplant participants. All area under the curve (AUC) images were normalized by the kidney pyruvate AUC. Due to the low signal level of bicarbonate, all bicarbonate images are shown with 5-time signal amplified.

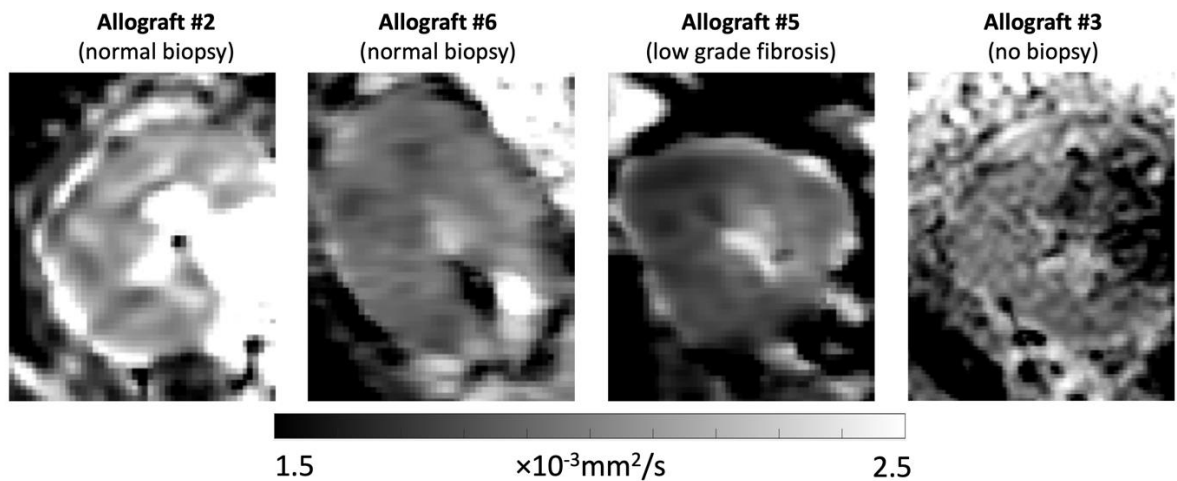

Figure S2. Apparent diffusion coefficient (ADC) maps from four kidney allografts.

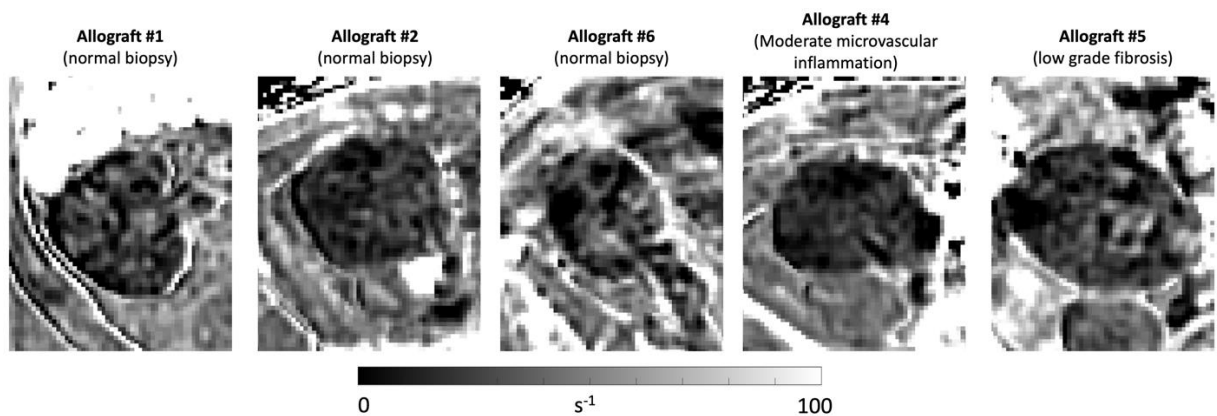

Figure S3. $R_2^*$ maps from five kidney allografts.